\pdfoutput=1
\documentclass[12pt,a4paper]{article}

\usepackage{ifthen} 
\newboolean{pdflatex}
\setboolean{pdflatex}{true} 

\newboolean{articletitles}
\setboolean{articletitles}{true} 

\newboolean{uprightparticles}
\setboolean{uprightparticles}{false} 

\usepackage{longtable} 
\usepackage{rotating}

\def\paperauthors{LHCb collaboration} 
\def\paperasciititle{Search for the rare decay of charmed baryon LambdaC into pmumu final state} 
\def\papertitle{Search for the rare decay of charmed baryon \Lc into \pmumu final state} 
\def\paperkeywords{{High Energy Physics}, {LHCb}} 
\def\papercopyright{\the\year\ CERN for the benefit of the LHCb collaboration} 
\def\paperlicence{CC BY 4.0 licence}
\def\paperlicenceurl{https://creativecommons.org/licenses/by/4.0/}

\usepackage[top=1in, bottom=1.25in, left=1in, right=1in]{geometry}

\usepackage{microtype}
\usepackage{lineno}  
\usepackage{xspace} 
\usepackage{caption} 

\usepackage{graphicx}  
\usepackage{color}
\usepackage{colortbl}
\graphicspath{{./figs/}} 

\usepackage{amsmath} 
\usepackage{amssymb}
\usepackage{amsfonts}
\usepackage{upgreek} 

\newcommand*\patchAmsMathEnvironmentForLineno[1]{%
\expandafter\let\csname old#1\expandafter\endcsname\csname #1\endcsname
\expandafter\let\csname oldend#1\expandafter\endcsname\csname
end#1\endcsname
 \renewenvironment{#1}%
   {\linenomath\csname old#1\endcsname}%
   {\csname oldend#1\endcsname\endlinenomath}%
}
\newcommand*\patchBothAmsMathEnvironmentsForLineno[1]{%
  \patchAmsMathEnvironmentForLineno{#1}%
  \patchAmsMathEnvironmentForLineno{#1*}%
}
\AtBeginDocument{%
\patchBothAmsMathEnvironmentsForLineno{equation}%
\patchBothAmsMathEnvironmentsForLineno{align}%
\patchBothAmsMathEnvironmentsForLineno{flalign}%
\patchBothAmsMathEnvironmentsForLineno{alignat}%
\patchBothAmsMathEnvironmentsForLineno{gather}%
\patchBothAmsMathEnvironmentsForLineno{multline}%
\patchBothAmsMathEnvironmentsForLineno{eqnarray}%
}

\usepackage{hyperxmp}

\usepackage[pdftex,
            pdfauthor={\paperauthors},
            pdftitle={\paperasciititle},
            pdfkeywords={\paperkeywords},
            pdfcopyright={Copyright (C) \papercopyright},
            pdflicenseurl={\paperlicenceurl}]{hyperref}

\usepackage[bottom,flushmargin,hang,multiple]{footmisc}

\usepackage[all]{hypcap} 

\usepackage{xspace} 
\usepackage{upgreek}

\def\lhcb   {\mbox{LHCb}\xspace}

\def\babar  {\mbox{BaBar}\xspace}

\def\MagUp {\mbox{\em Mag\kern -0.05em Up}\xspace}

\ifthenelse{\boolean{uprightparticles}}%
{

 \def\Peta        {\ensuremath{\upeta}\xspace}

 \def\Pmu         {\ensuremath{\upmu}\xspace}

 \def\Prho        {\ensuremath{\uprho}\xspace}

 \def\Pphi        {\ensuremath{\upphi}\xspace}

 \def\Pomega      {\ensuremath{\upomega}\xspace}                 

 \def\PDelta      {\ensuremath{\Delta}\xspace}                 
 \def\PXi         {\ensuremath{\Xi}\xspace}                 
 \def\PLambda     {\ensuremath{\Lambda}\xspace}                 
 \def\PSigma      {\ensuremath{\Sigma}\xspace}                 
 \def\POmega      {\ensuremath{\Omega}\xspace}                 
 \def\PUpsilon    {\ensuremath{\Upsilon}\xspace}
 \let\oldPi\Pi
 \def\PPi         {\ensuremath{\oldPi}\xspace}

 \def\PB      {\ensuremath{\mathrm{B}}\xspace}                 
                  
 \def\PD      {\ensuremath{\mathrm{D}}\xspace}

 \def\PK      {\ensuremath{\mathrm{K}}\xspace}

 \def\Pb      {\ensuremath{\mathrm{b}}\xspace}                 
 \def\Pc      {\ensuremath{\mathrm{c}}\xspace}

 \def\Pi      {\ensuremath{\mathrm{i}}\xspace}

 \def\Pp      {\ensuremath{\mathrm{p}}\xspace}

 \def\Ps      {\ensuremath{\mathrm{s}}\xspace}

 \def\thebaroffset{0.0em}
}
{

 \def\Peta        {\ensuremath{\eta}\xspace}

 \def\Pmu         {\ensuremath{\mu}\xspace}

 \def\Prho        {\ensuremath{\rho}\xspace}

 \def\Pphi        {\ensuremath{\phi}\xspace}

 \def\Pomega      {\ensuremath{\omega}\xspace}                 
 \mathchardef\PDelta="7101
 \mathchardef\PXi="7104
 \mathchardef\PLambda="7103
 \mathchardef\PSigma="7106
 \mathchardef\POmega="710A
 \mathchardef\PUpsilon="7107
 \mathchardef\PPi="7105
                  
 \def\PB      {\ensuremath{B}\xspace}                 
                  
 \def\PD      {\ensuremath{D}\xspace}

 \def\PK      {\ensuremath{K}\xspace}

 \def\Pb      {\ensuremath{b}\xspace}                 
 \def\Pc      {\ensuremath{c}\xspace}

 \def\Pi      {\ensuremath{i}\xspace}

 \def\Pp      {\ensuremath{p}\xspace}

 \def\Ps      {\ensuremath{s}\xspace}

 \def\thebaroffset{0.18em}
}
\newcommand{\offsetoverline}[2][\thebaroffset]{\kern #1\overline{\kern -#1 #2}}%

\makeatletter
\ifcase \@ptsize \relax
  \newcommand{\miniscule}{\@setfontsize\miniscule{4}{5}}
\or
  \newcommand{\miniscule}{\@setfontsize\miniscule{5}{6}}
\or
  \newcommand{\miniscule}{\@setfontsize\miniscule{5}{6}}
\fi
\makeatother

\DeclareRobustCommand{\optbar}[1]{\shortstack{{\miniscule (\rule[.5ex]{1.25em}{.18mm})}
  \\ [-.7ex] $#1$}}

\def\mup        {{\ensuremath{\Pmu^+}}\xspace}
\def\mun        {{\ensuremath{\Pmu^-}}\xspace} 

\def\mumu       {{\ensuremath{\Pmu^+\Pmu^-}}\xspace}

\def\squark    {{\ensuremath{\Ps}}\xspace}

\def\cquark    {{\ensuremath{\Pc}}\xspace}

\def\bquark    {{\ensuremath{\Pb}}\xspace}

\def\KorKbar {\kern \thebaroffset\optbar{\kern -\thebaroffset \PK}{}\xspace}

\def\D       {{\ensuremath{\PD}}\xspace}

\def\DorDbar {\kern \thebaroffset\optbar{\kern -\thebaroffset \PD}\xspace}

\def\Dp      {{\ensuremath{\D^+}}\xspace}
\def\Dm      {{\ensuremath{\D^-}}\xspace}

\def\DpDm    {\ensuremath{\Dp {\kern -0.16em \Dm}}\xspace}

\def\B       {{\ensuremath{\PB}}\xspace}

\def\BorBbar {\kern \thebaroffset\optbar{\kern -\thebaroffset \PB}\xspace}

\def\Bd      {{\ensuremath{\B^0}}\xspace}

\def\BdorBdbar {\kern \thebaroffset\optbar{\kern -\thebaroffset \Bd}\xspace}

\def\Bs      {{\ensuremath{\B^0_\squark}}\xspace}

\def\BsorBsbar {\kern \thebaroffset\optbar{\kern -\thebaroffset \Bs}\xspace}

\def\Y#1S{\ensuremath{\PUpsilon{(#1S)}}\xspace}

\def\proton      {{\ensuremath{\Pp}}\xspace}

\def\Lz          {{\ensuremath{\PLambda}}\xspace}

\def\LorLbar     {\kern \thebaroffset\optbar{\kern -\thebaroffset \PLambda}\xspace}

\def\Lc          {{\ensuremath{\Lz^+_\cquark}}\xspace}

\newcommand{\decay}[2]{\ensuremath{#1\!\to #2}\xspace} 

\def\to                 {\ensuremath{\rightarrow}\xspace}

\def\AT#1     {\ensuremath{A_{\mathrm{T}}^{#1}}\xspace}           

\def\C#1      {\ensuremath{\mathcal{C}_{#1}}\xspace}                       
\def\Cp#1     {\ensuremath{\mathcal{C}_{#1}^{'}}\xspace}                    
\def\Ceff#1   {\ensuremath{\mathcal{C}_{#1}^{\mathrm{(eff)}}}\xspace}        
\def\Cpeff#1  {\ensuremath{\mathcal{C}_{#1}^{'\mathrm{(eff)}}}\xspace}       
\def\Ope#1    {\ensuremath{\mathcal{O}_{#1}}\xspace}                       
\def\Opep#1   {\ensuremath{\mathcal{O}_{#1}^{'}}\xspace}                    

\newcommand{\nospaceunit}[1]{\ensuremath{\text{#1}}}       
\newcommand{\aunit}[1]{\ensuremath{\text{\,#1}}}       

\newcommand{\tev}{\aunit{Te\kern -0.1em V}\xspace}
\newcommand{\gev}{\aunit{Ge\kern -0.1em V}\xspace}
\newcommand{\mev}{\aunit{Me\kern -0.1em V}\xspace}
\newcommand{\kev}{\aunit{ke\kern -0.1em V}\xspace}
\newcommand{\ev}{\aunit{e\kern -0.1em V}\xspace}
 
\newcommand{\mevc}{\ensuremath{\aunit{Me\kern -0.1em V\!/}c}\xspace}
\newcommand{\gevc}{\ensuremath{\aunit{Ge\kern -0.1em V\!/}c}\xspace}
\newcommand{\mevcc}{\ensuremath{\aunit{Me\kern -0.1em V\!/}c^2}\xspace}
\newcommand{\gevcc}{\ensuremath{\aunit{Ge\kern -0.1em V\!/}c^2}\xspace}

\def\mum  {\ensuremath{\,\upmu\nospaceunit{m}}\xspace}

\def\fb   {\ensuremath{\aunit{fb}}\xspace}
\def\invfb   {\ensuremath{\fb^{-1}}\xspace}

\newcommand{\chisq}{\ensuremath{\chi^2}\xspace}

\newcommand{\chisqip}{\ensuremath{\chi^2_{\text{IP}}}\xspace}

\def\gsim{{~\raise.15em\hbox{$>$}\kern-.85em
          \lower.35em\hbox{$\sim$}~}\xspace}
\def\lsim{{~\raise.15em\hbox{$<$}\kern-.85em
          \lower.35em\hbox{$\sim$}~}\xspace}

\def\pt         {\ensuremath{p_{\mathrm{T}}}\xspace}

\def\ptot       {\ensuremath{p}\xspace}

\def\evtgen     {\mbox{\textsc{EvtGen}}\xspace}

\def\geant      {\mbox{\textsc{Geant4}}\xspace}

\def\photos     {\mbox{\textsc{Photos}}\xspace}

\def\pythia     {\mbox{\textsc{Pythia}}\xspace}

\def\tell1  {TELL1\xspace}
\def\ukl1   {UKL1\xspace}

\newcommand{\lhcborcid}[1]{\href{https://orcid.org/#1}{\hspace*{0.1em}\raisebox{-0.45ex}{\includegraphics[width=1em]{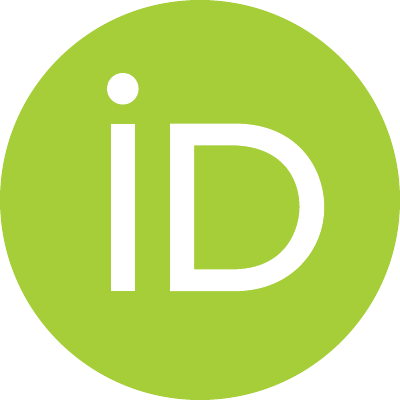}}}}

\usepackage{cite} 
\usepackage{mciteplus}

\def\LcToppipi  {\decay{\Lc}{\proton \pi^+\pi^-}}
\def\LcTopmumu  {\decay{\Lc}{\proton \mumu}}
 
\def\LcTopphi  {\decay{\Lc}{\proton \Pphi (\to \mumu)}}
\def\LcTopomega  {\decay{\Lc}{\proton \Pomega (\to \mumu)}}
\def\LcTopeta  {\decay{\Lc}{\proton \eta (\to \mumu)}}

\def\LcToprho  {\decay{\Lc}{\proton \Prho (\to \mumu)}}

\newcommand{\BRof}[1]{\ensuremath{{\cal B}(#1)}\xspace}

\def\pmumu  {\ensuremath{\proton \mup \mun}\xspace}

\def\Lcpphino  {\decay{\Lc}{\proton \Pphi}}
\def\Lcpomegano  {\decay{\Lc}{\proton \Pomega}}
\def\Lcprhono  {\decay{\Lc}{\proton \Prho}}
\def\Lcpetano  {\decay{\Lc}{\proton \Peta}}
\def\LcToPmm {\decay{\Lc}{\proton\mup\mun}}
\def\bTosll {\decay{\bquark}{\squark\ell\ell}}
\newcommand{\signal}{\ensuremath{signal}\xspace}
\def\CLs {{CL$_{\rm s}$}\xspace}
\def\etamumu {\decay{\Peta}{\mup\mun}}
\def\rhomumu {\decay{\Prho}{\mup\mun}}
\def\phimumu  {\decay{\Pphi}{\mumu}}
\def\omegamumu  {\decay{\Pomega}{\mumu}}

\def\mumum       {{\ensuremath{m_{\mu\mu}}}\xspace}
\def\lowq {{\ensuremath{low\text{-}m}}\xspace}
\def\highq {{\ensuremath{high\text{-}m}}\xspace}

\begin{document}

\renewcommand{\thefootnote}{\fnsymbol{footnote}}
\setcounter{footnote}{1}


\begin{titlepage}
\pagenumbering{roman}

\vspace*{-1.5cm}
\centerline{\large EUROPEAN ORGANIZATION FOR NUCLEAR RESEARCH (CERN)}
\vspace*{1.5cm}
\noindent
\begin{tabular*}{\linewidth}{lc@{\extracolsep{\fill}}r@{\extracolsep{0pt}}}
\ifthenelse{\boolean{pdflatex}}
{\vspace*{-1.5cm}\mbox{\!\!\!\includegraphics[width=.14\textwidth]{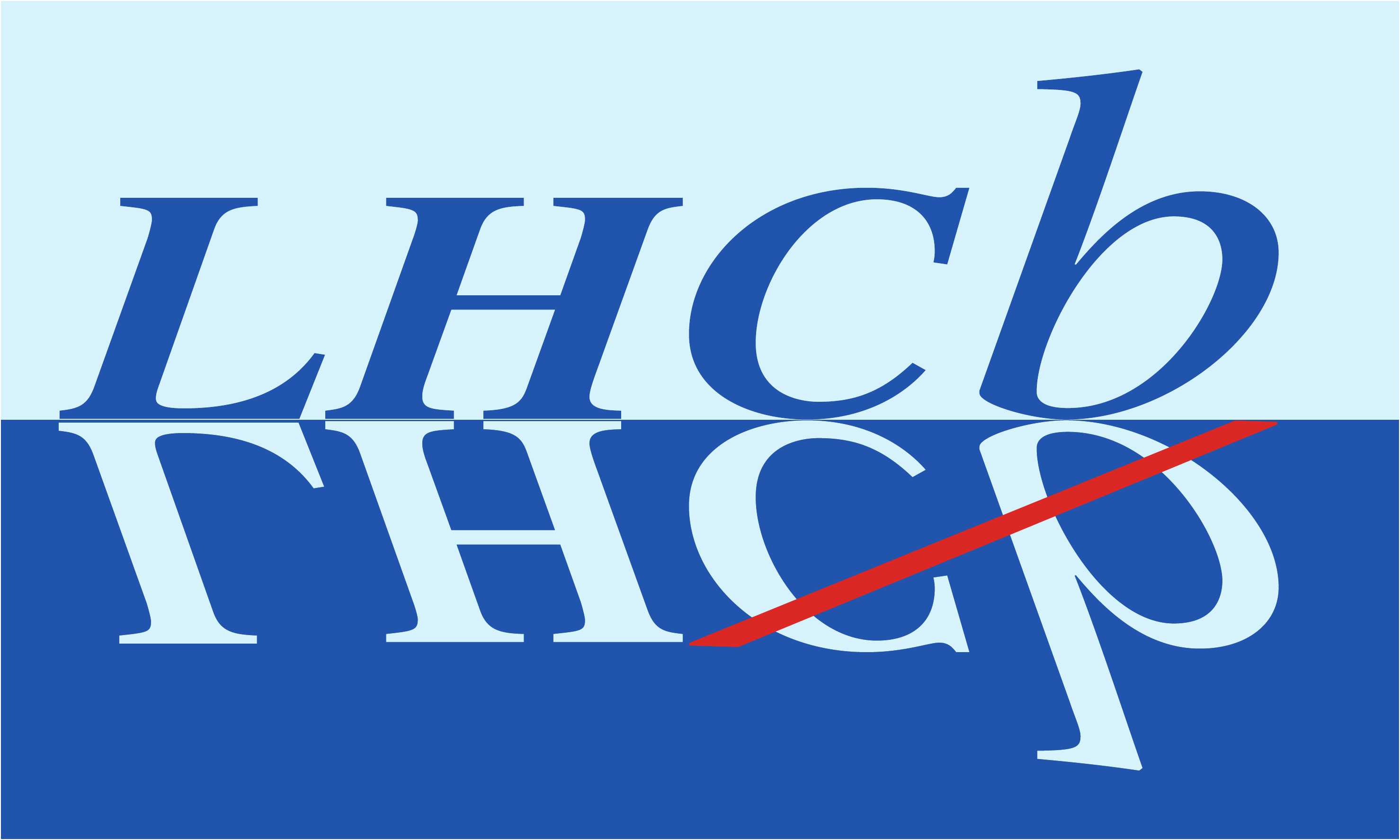}} & &}%
{\vspace*{-1.2cm}\mbox{\!\!\!\includegraphics[width=.12\textwidth]{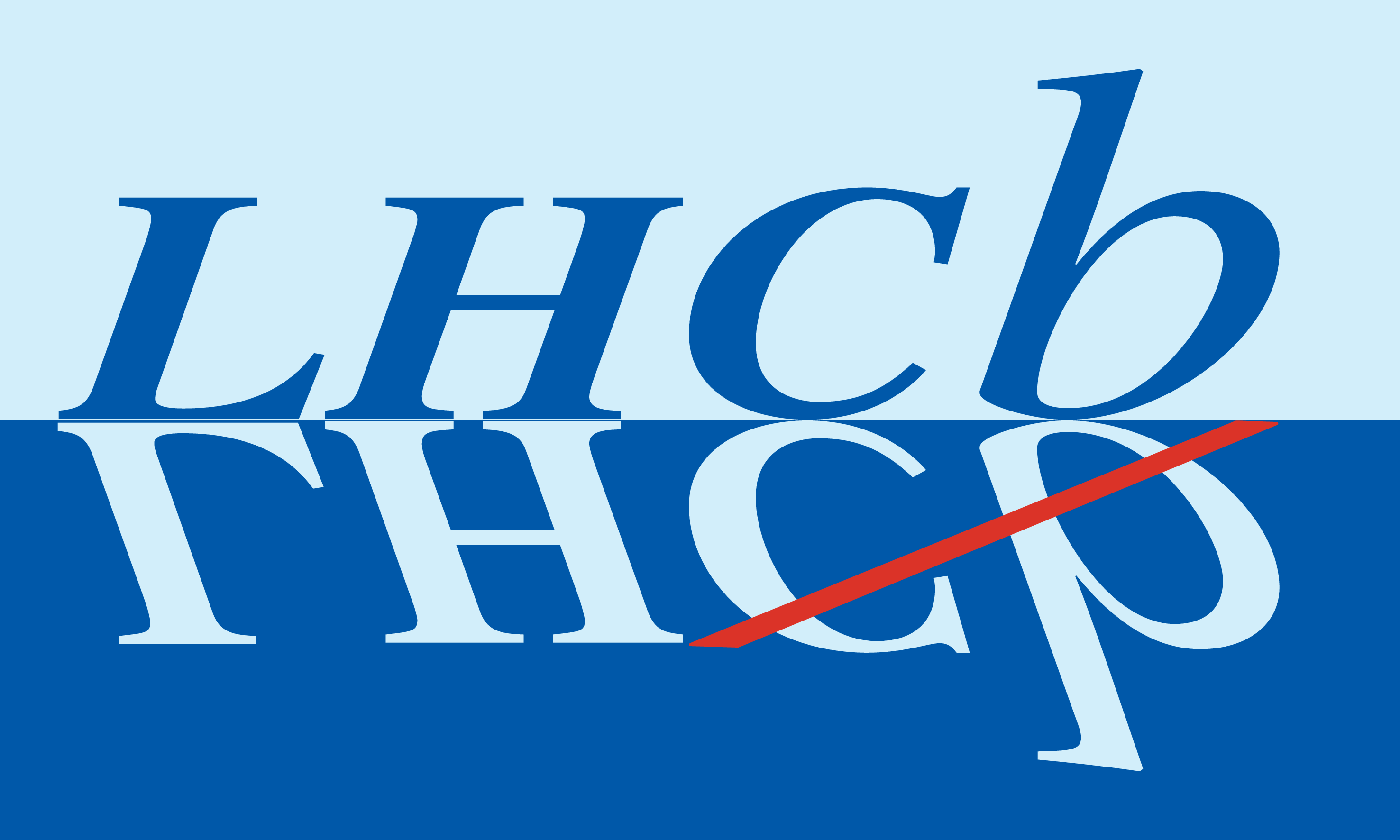}} & &}%
\\
 & & CERN-EP-2024-158 \\  
 & & LHCb-PAPER-2024-005 \\  
 & & September 26, 2024 \\ 
 & & \\
\end{tabular*}

\vspace*{4.0cm}

{\normalfont\bfseries\boldmath\huge
\begin{center}
  \papertitle 
\end{center}
}

\vspace*{2.0cm}

\begin{center}
\paperauthors\footnote{Authors are listed at the end of this paper.}
\end{center}

\vspace{\fill}

\begin{abstract}
  \noindent
A search for the nonresonant \LcTopmumu decay is performed using proton-proton collision data recorded at a centre-of-mass energy of $13\,\tev$ by the LHCb experiment,  corresponding to an integrated luminosity of $5.4\invfb$. No evidence for the decay is found in the dimuon invariant-mass regions where the expected contributions of resonances is subdominant. 
The upper limit on the branching fraction of the \LcTopmumu decay is determined to be \mbox{$2.9~(3.2) \times10^{-8}~{\rm at}~90\%~(95\%)$}~confidence level. The branching fractions in the dimuon invariant-mass regions dominated by the $\eta$, $\rho$ and $\omega$ resonances are also determined. 
\end{abstract}

\vspace*{2.0cm}

\begin{center}
  Published in Phys.~Rev.~D 110 (2024) 052007 
\end{center}

\vspace{\fill}

{\footnotesize 
\centerline{\copyright~\papercopyright. \href{\paperlicenceurl}{\paperlicence}.}}
\vspace*{2mm}

\end{titlepage}


\newpage
\setcounter{page}{2}
\mbox{~}


\renewcommand{\thefootnote}{\arabic{footnote}}
\setcounter{footnote}{0}

\cleardoublepage


\pagestyle{plain} 
\setcounter{page}{1}
\pagenumbering{arabic}


\section{Introduction}
\label{sec:Introduction}

The flavour-changing neutral-current decay\footnote{The charge-conjugate mode is implied throughout the paper.} \LcToPmm is  heavily suppressed in the Standard Model (SM) as it is forbidden at tree level and is affected by the Glashow--Iliopoulos--Maiani (GIM) mechanism~\cite{GIM}. 
The phenomenology of the decay is discussed in Ref.~\cite{Meinel:2017ggx} and the SM predictions for $q^2$ differential branching fractions, where $q^2$ is the dimuon invariant-mass squared, are calculated in Ref.~\cite{GoltzRare2021}. The branching fraction ($\mathcal{B}$) associated with the short-distance contributions to the $c\to u \mu^+ \mu^-$ transition
is expected to be below $10^{-8}$ in the SM, with an enhancement up to $10^{-6}$ due to long-distance processes involving intermediate vector-meson resonances that decay 
to a lepton pair~\cite{CHARM_RARE_FAJFER,CHARM_RARE_BURDMAN}.
The estimation of resonant SM contributions relies on the measured values for the corresponding branching fractions and the isospin relation between $\rho$
and $\omega$ states, to overcome the lack of experimental information on $\mathcal{B}$(\Lcprhono). Strong phases between the resonances, and thus the effect of interference, are \textit{a priori} unknown. 
 
Interesting deviations from the SM predictions have been observed in the equivalent type of transition for the $\bquark$ quark, \bTosll,  
in particular in angular observables~\cite{LHCb-PAPER-2020-041}. These anomalies motivate  the extension of this physics program into the charm sector, which is far less explored both experimentally and theoretically. The first results, such as an angular analysis of charm mesons~\cite{LHCb-PAPER-2021-035}, are already available. A discussion of the discovery potential of effects beyond the SM in rare charm-baryon decays can be found in Ref.~\cite{MitzelRare2021}. 

The LHCb collaboration previously searched for the \LcToPmm decay using the proton-proton ($pp$) data collected in 2011--2012~\cite{LHCb-PAPER-2017-039} and placed an upper limit of $7.7\times 10^{-8}$ at 90\% confidence level (CL), thereby improving the result from the \babar experiment~\cite{BABAR} by two orders of magnitude. The analysis of data collected in 2016--2018 presented in this paper follows the methodology of the previous analysis, and relies on improved techniques with the use of the \CLs method~\cite{CLs} implementation based on unbinned likelihoods, the incorporation of background due to particle misidentification into the data model, a different definition of the short-distance signal and a more detailed treatment of the entire range of dimuon invariant mass where resonances are expected to dominate. Accordingly, the new results for the short distance contribution should not be combined with the results based on the 2011--2012 data.

The measurement is carried out in different regions of dimuon invariant mass \mumum and the \LcTopphi decay is used for normalisation since it has the same final state as the signal and is relatively abundant.

\section{Detector and simulation}
\label{sec:Detector}
The \lhcb detector~\cite{LHCb-DP-2008-001,LHCb-DP-2014-002} is a single-arm forward
spectrometer covering the \mbox{pseudorapidity} range $2<\eta <5$,
designed for the study of particles containing \bquark or \cquark
quarks. The detector includes a high-precision tracking system
consisting of a silicon-strip vertex detector surrounding the $pp$
interaction region~\cite{LHCb-DP-2014-001}, a large-area silicon-strip detector located
upstream of a dipole magnet with a bending power of about
$4{\mathrm{\,T\, m}}$, 
and three stations of silicon-strip detectors and straw
drift tubes~\cite{LHCb-DP-2017-001}
placed downstream of the magnet.
The tracking system provides a measurement of the momentum, \ptot, of charged particles with
a relative uncertainty that varies from 0.5\% at low momentum to 1.0\% at 200\gevc.
The minimum distance of a track to a primary $pp$ collision vertex (PV), the impact parameter (IP), 
is measured with a resolution of $(15+29/\pt)\mum$,
where \pt is the component of the momentum transverse to the beam, in\,\gevc.
Different types of charged hadrons are distinguished using information
from two ring-imaging Cherenkov detectors~\cite{LHCb-DP-2012-003}. 
Photons, electrons and hadrons are identified by a calorimeter system consisting of
scintillating-pad and preshower detectors, an electromagnetic
and a hadronic calorimeter. Muons are identified by a
system composed of alternating layers of iron and multiwire
proportional chambers~\cite{LHCb-DP-2012-002}.
The online event selection is performed by a trigger system~\cite{LHCb-DP-2012-004}, 
which consists of a hardware stage, based on information from the calorimeter and muon
systems, followed by a software stage, which applies a full event
reconstruction.

Simulated samples of \Lc baryons are produced to model the effects of the detector acceptance and the imposed selection requirements.
The $pp$ collisions are generated using
\pythia~\cite{Sjostrand:2007gs,*Sjostrand:2006za} 
with a specific \lhcb configuration~\cite{LHCb-PROC-2010-056}. 
Decays of unstable particles are described by \evtgen~\cite{Lange:2001uf}, in which final state radiation is generated using \photos~\cite{davidson2015photos}. In the case of \LcTopmumu decays, a model with final state particles equally distributed in the phase space is used.
The interaction of the generated particles with the detector, and its response,
is implemented using the \geant
toolkit~\cite{Allison:2006ve, *Agostinelli:2002hh} as described in
Ref.~\cite{LHCb-PROC-2011-006}. 
The underlying $pp$ interaction and \Lc baryon are reused multiple times to speed up the simulation process, with independently generated \Lc decay products~\cite{LHCb-DP-2018-004}. 

\section{Candidate selection}
The selection criteria largely follow those described in Ref.~\cite{LHCb-PAPER-2017-039}. The \Lc candidates are reconstructed by combining a pair of oppositely charged tracks identified as muons with an additional track identified as a proton. The three tracks are required to be compatible with originating from a common vertex and have significant \chisqip with respect to all PVs. Here, \chisqip is defined as the difference in the vertex-fit \chisq of a given PV, reconstructed with and without the track under consideration. The \Lc candidates must have a decay vertex displaced from any PV and be compatible with originating from one of the PVs. 

The background formed by random combinations of tracks, referred to as combinatorial background, is suppressed using a multivariate classifier. A boosted decision tree (BDT) algorithm~\cite{Breiman}, as implemented in the TMVA toolkit~\cite{Hocker:2007ht,TMVA4}, is trained using supervised learning with tenfold cross-validation~\cite{fold} to achieve an unbiased classifier response.
The BDT discrimination is done in two steps. A first BDT classifier employs topological and kinematic variables of the \Lc candidates. 
A relatively loose requirement on the response of that BDT is applied as the last preselection step to enhance the purity of the normalisation data sample. 
After preselection, simulated samples of \LcTopphi decays are compared with data, and weights are determined to improve the agreement between data and simulation for the \Lc transverse momentum and the number of tracks in the event.
A second BDT classifier is trained  using a set of variables that includes those of the first classifier and, in addition, kinematic, vertex and isolation variables related to the decay products.
For the two classifiers, simulated events containing \LcTopmumu decays are used as a proxy for the signal, while the background is modelled using sideband events with a $p\mup\mun$ invariant mass within [2140, 2251] and [2321, 2430]\,\mevcc.
 The optimisation of final requirements is performed in three dimensions: the response value of the second BDT classifier and two variables characterising the proton and muon particle identification (PID). In this optimisation, the simulation is corrected for imperfections in modeling of the PID performance~\cite{LHCb-DP-2018-001}. 
The final requirements are determined using pseudoexperiments by finding the best expected upper limit (using the \CLs method) on the ratio of branching fractions of the signal relative to the normalisation channel. 
At this stage, only two background sources are expected to significantly contribute: combinatorial background and \LcToppipi decays where both pions are misidentified as muons. In each pseudoexperiment, pseudodata for the signal and these two background components are drawn from the three-dimensional probability density functions determined after the preselection stage. 

Six dimuon invariant-mass ranges are considered as presented in Table~\ref{tab:mumudef}. The nonresonant signal decay is searched for in two \mumum regions, \lowq and \highq, where the expected contributions of resonances are subdominant. In the following, the combination of these two ranges is referred to as the \emph{signal} region. 
The non \emph{signal} range is divided into regions where either the $\eta$, $\rho$, $\omega$ or $\phi$ resonance dominates.
The branching fraction results quoted for a given resonant region are labeled by the resonant decay expected to dominate in that region. However, due to the relatively low sample size, no attempt to separate the resonant contributions within a given dimuon invariant-mass range is made. Therefore it has to be emphasised that the results must be interpreted as corresponding to that particular \mumum region rather than to a specific resonant decay.
\begin{table}[tb]
\caption{Definitions of the dimuon invariant-mass ranges. The lower (upper) bound of the \lowq (\highq) region corresponds to the phase-space limits. 
}
\label{tab:mumudef}
\begingroup
\setlength{\tabcolsep}{10pt} 
\renewcommand{\arraystretch}{1.1} 
\begin{center}

\begin{tabular}{lc}
\hline
Region & Range [\mevcc] \\ 
\hline
\lowq     &  $~~~211.32 < m_{\mumu} < ~507.86$    \\ 
\highq    &  $~~1059.46 < m_{\mumu} < 1348.13$    \\ 
$\eta$    &  $~~~507.86 < m_{\mumu} < ~587.86$    \\ 
$\omega$  &  $~~~742.65 < m_{\mumu} < ~822.65$    \\ 
$\rho$    &  $~~~587.86 < m_{\mumu} < ~742.65$    \\ 
                &  or $822.65~ < m_{\mumu} < ~965.20~$  \\ 
$\phi$    &  $~~~~979.46~ < m_{\mumu} < 1059.46$  \\
\hline
\end{tabular}
\end{center}
\endgroup
\end{table}
%

\section{Normalisation and invariant-mass fit}
\label{sec:fit}
The \LcTopmumu branching fraction is determined from the measured signal yield $N_{\rm sig}$  according to
\begin{equation}
{\BRof{\LcTopmumu}  = \alpha \times N_{\rm sig}},
\label{eq:normalisation}
\end{equation}
where the normalisation factor $\alpha$ is given by
\begin{equation}
{\alpha  = \frac{\rm {\epsilon\mathstrut_{norm}}}{\rm{\epsilon\mathstrut_{\rm sig}}\times \it{N}_{\rm norm }}}
\times
{\BRof{\Lcpphino} \times \BRof{\phimumu}}.
\label{eq:alpha}
\end{equation}
Here, $N_{\rm norm}$ is the yield of \LcTopphi decays, 
while $\rm\epsilon\mathstrut_{sig}$ and $\rm\epsilon\mathstrut_{norm}$ represent the corresponding total efficiencies for signal and normalisation channels, respectively. 
In the case of the ratio of branching fractions \BRof{\LcTopmumu}/\BRof{\LcTopphi}, the corresponding normalisation factor $\alpha$ takes the form 
${\rm \epsilon\mathstrut_{norm}}/({\rm \epsilon\mathstrut_{sig} \times \it{N}_{\rm norm} )}$. The ratio of efficiencies ${\rm \epsilon\mathstrut_{norm}}/{\rm \epsilon\mathstrut_{sig}}$ for the signal regions is compatible with unity within statistical uncertainty.

An extended unbinned maximum-likelihood fit to the \pmumu invariant mass is performed in each \mumum region, where
the signal shape of \Lc candidates is described by a Crystal Ball function~\cite{Skwarnicki:1986xj}. The background is described using two distinct components: the combinatorial background, modeled by an exponential shape, and the \LcToppipi peaking background, represented by the sum of a Crystal Ball function and a Gaussian function
The shapes of both the signal and the \LcToppipi background are determined from simulation for each \mumum region and fixed in the fit to data. Weights are applied to simulated \LcToppipi events to match the phase-space distribution observed in Ref.~\cite{LHCb-PAPER-2017-044}.

In each \mumum region, there are four parameters in the fit. Three of them are free parameters: the signal yield, the combinatorial background yield and the exponent describing the shape of the combinatorial background. The fourth one corresponds to the \LcToppipi background. Its total yield in all regions is a free parameter, and the fraction in each region is constrained based on simulation.

\begin{figure}
  \begin{center}
     \includegraphics[width=0.52\linewidth]{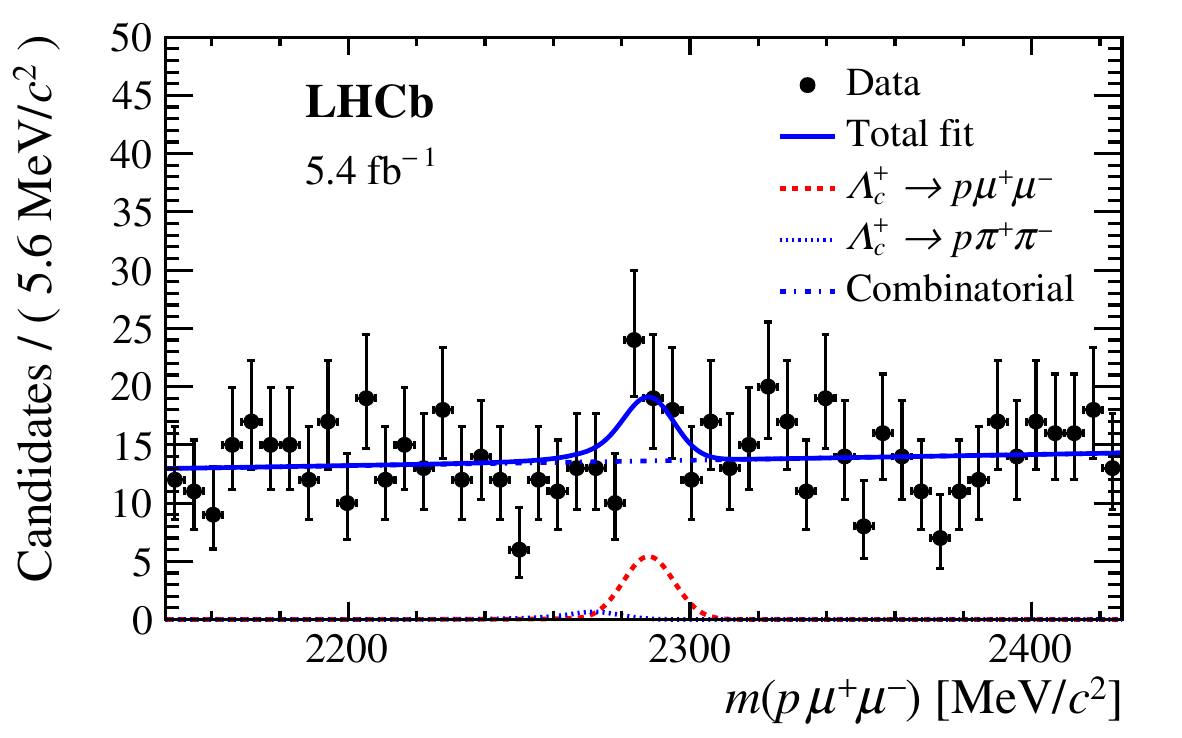}
     \includegraphics[width=0.49\linewidth]{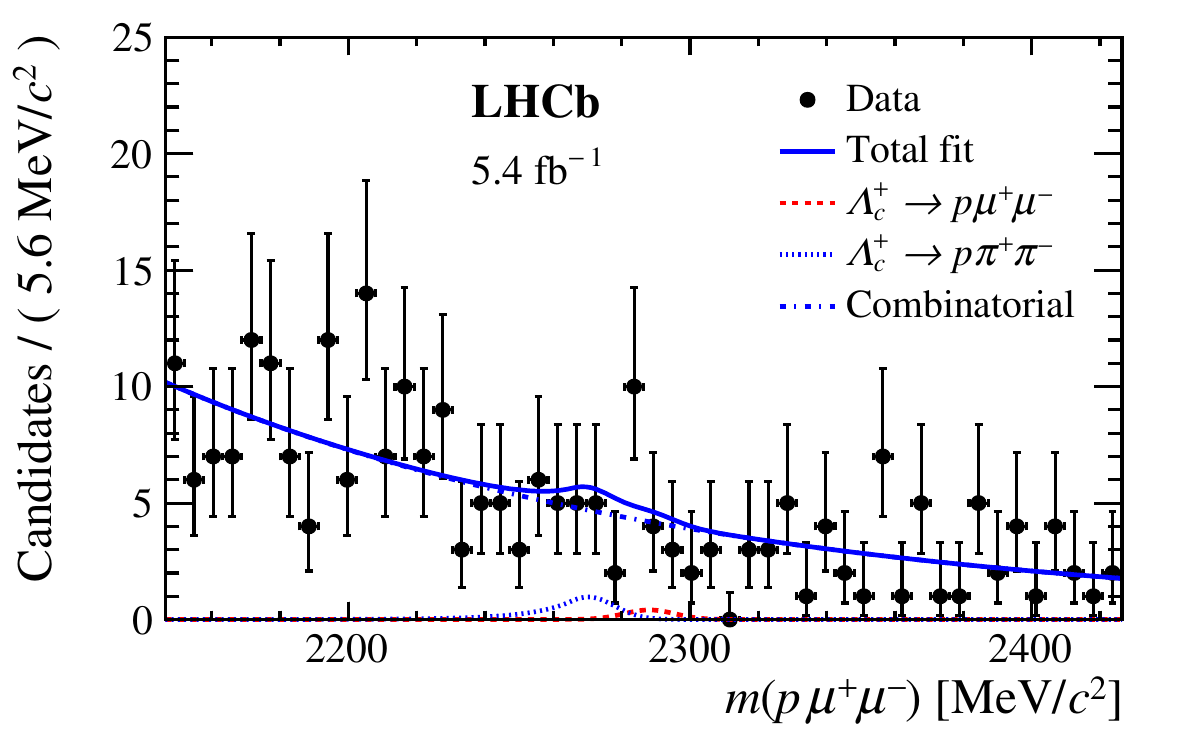}
     \includegraphics[width=0.49\linewidth]{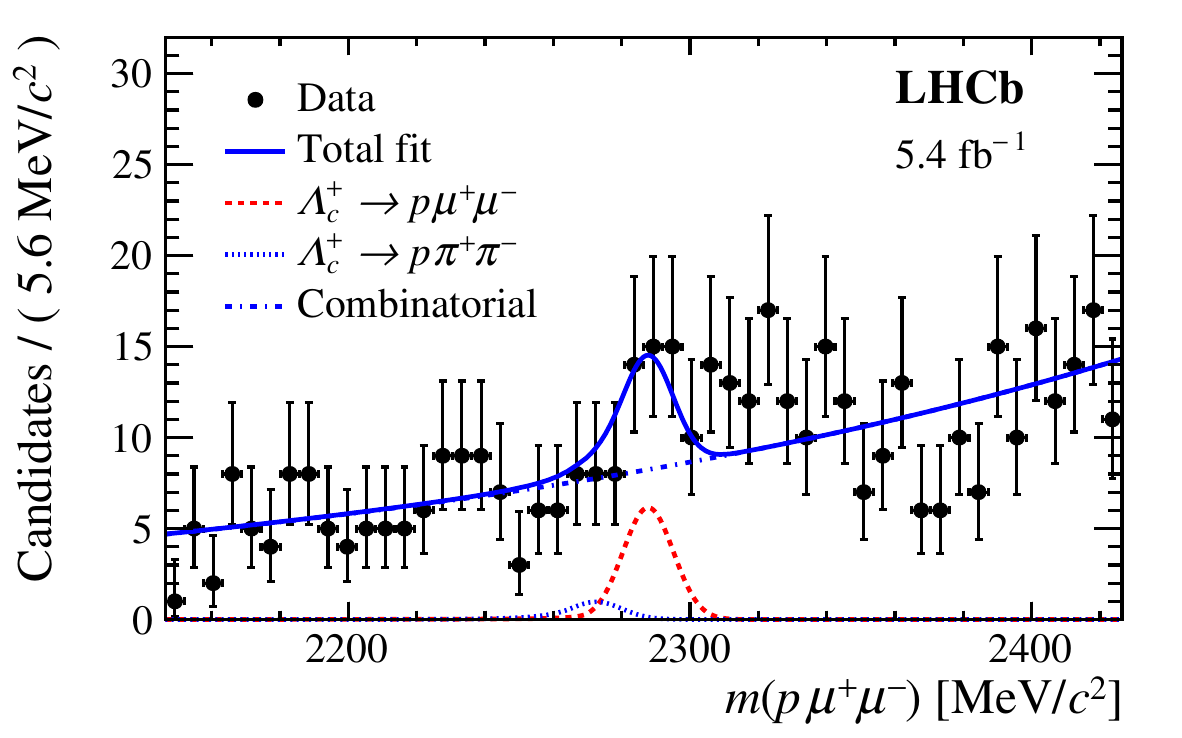}
     \vspace*{-0.6cm}
  \end{center}
  \caption{
    \small 
      Distributions of $p\mu^+\mu^-$ invariant mass for (top) \textit{signal}, (bottom left) $low$-$m$ and (bottom right) $high$-$m$ regions. The results of the fit are overlaid.
   }
   \label{fig:massfitsignal}
\end{figure}

\begin{figure}
  \begin{center}
     \includegraphics[width=0.49\linewidth]{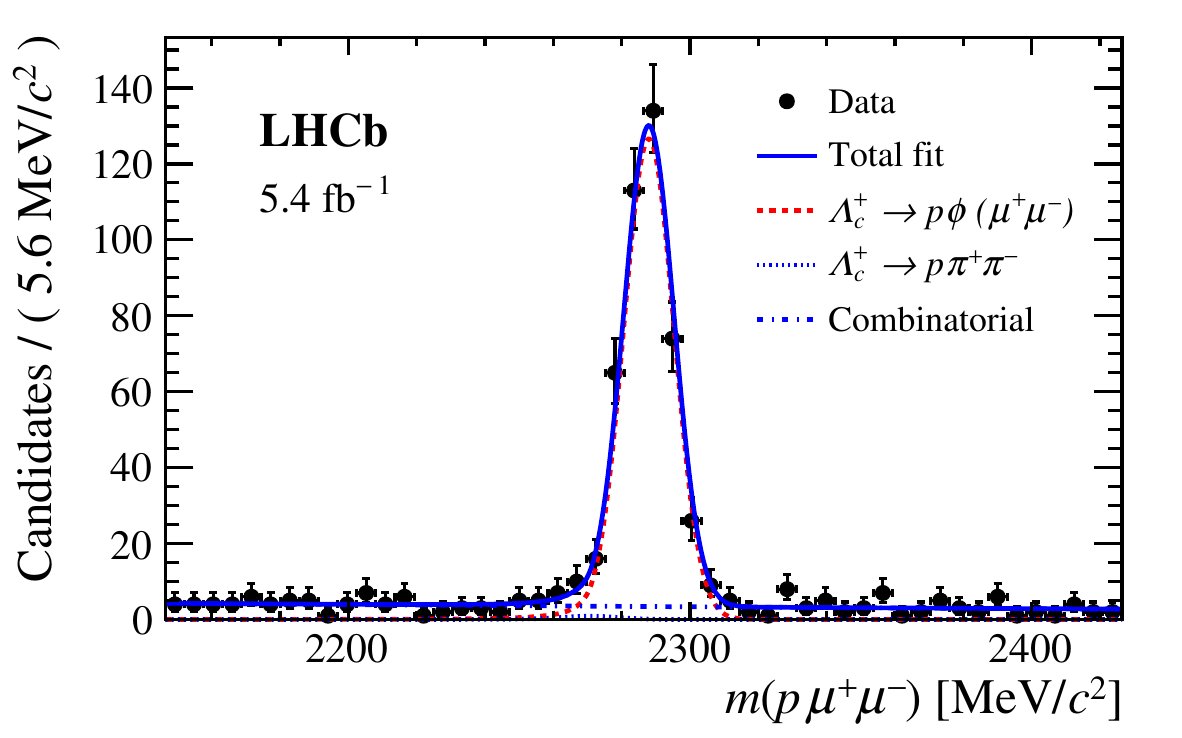}
     \includegraphics[width=0.49\linewidth]{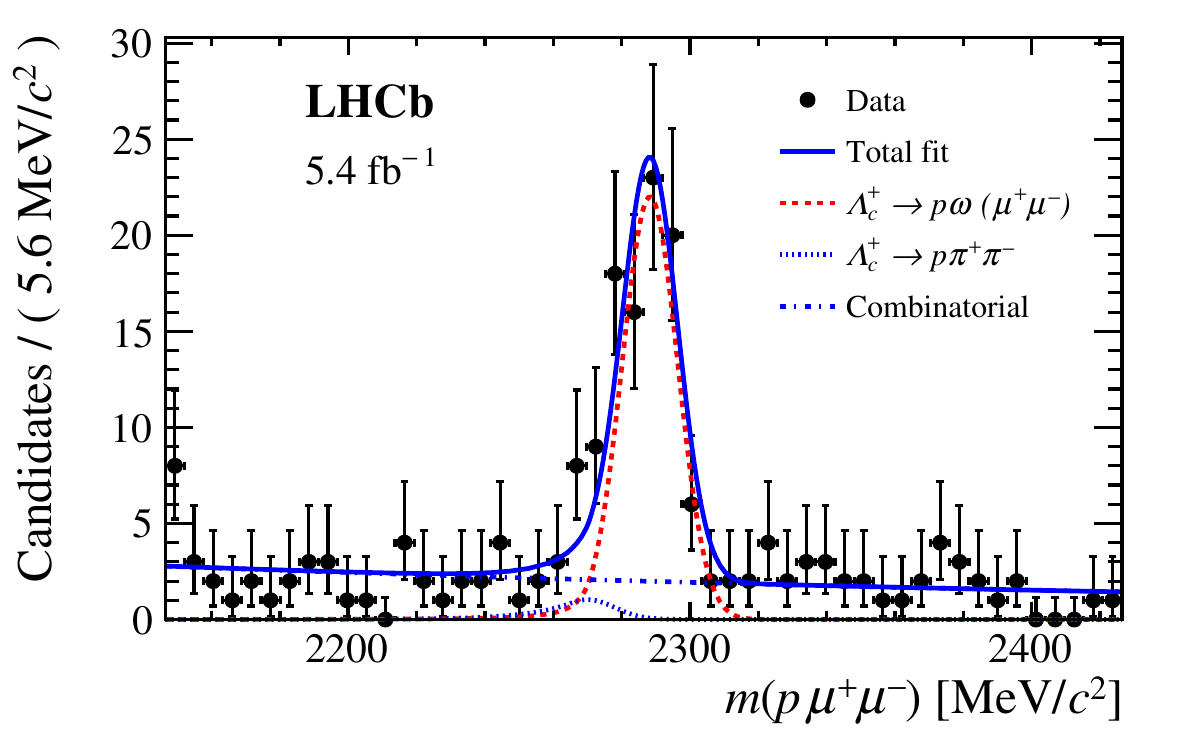}
     \includegraphics[width=0.49\linewidth]{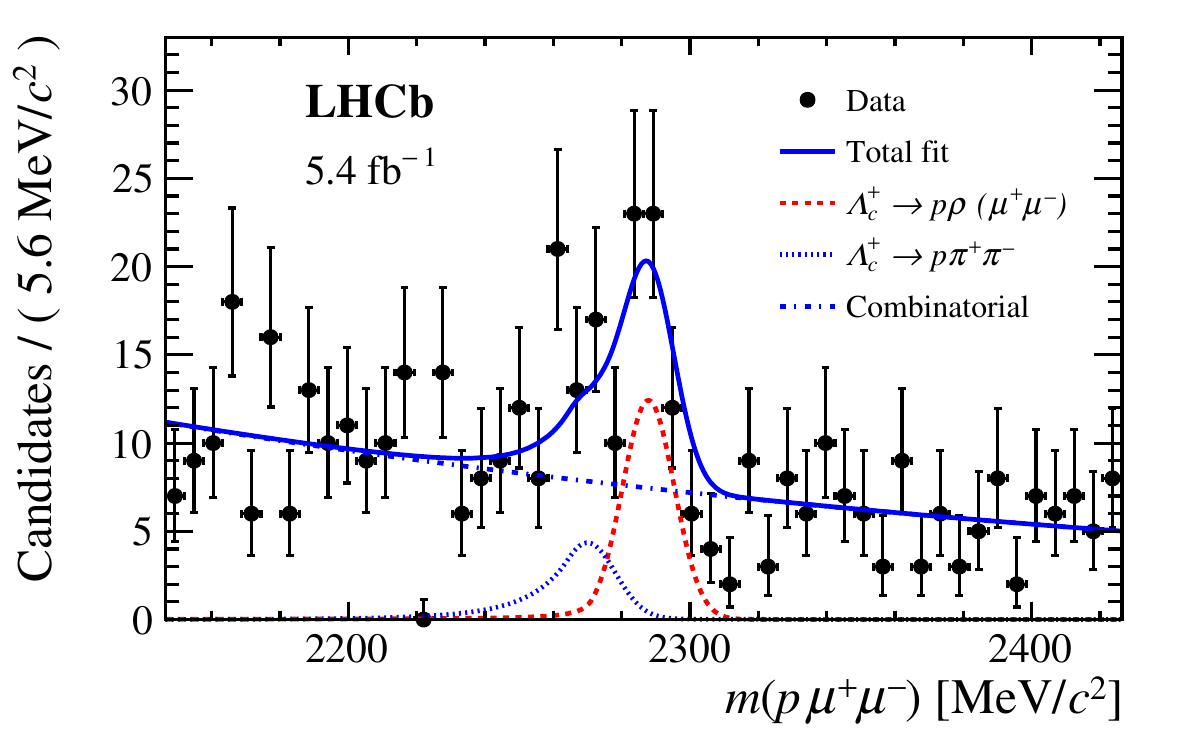}
     \includegraphics[width=0.49\linewidth]{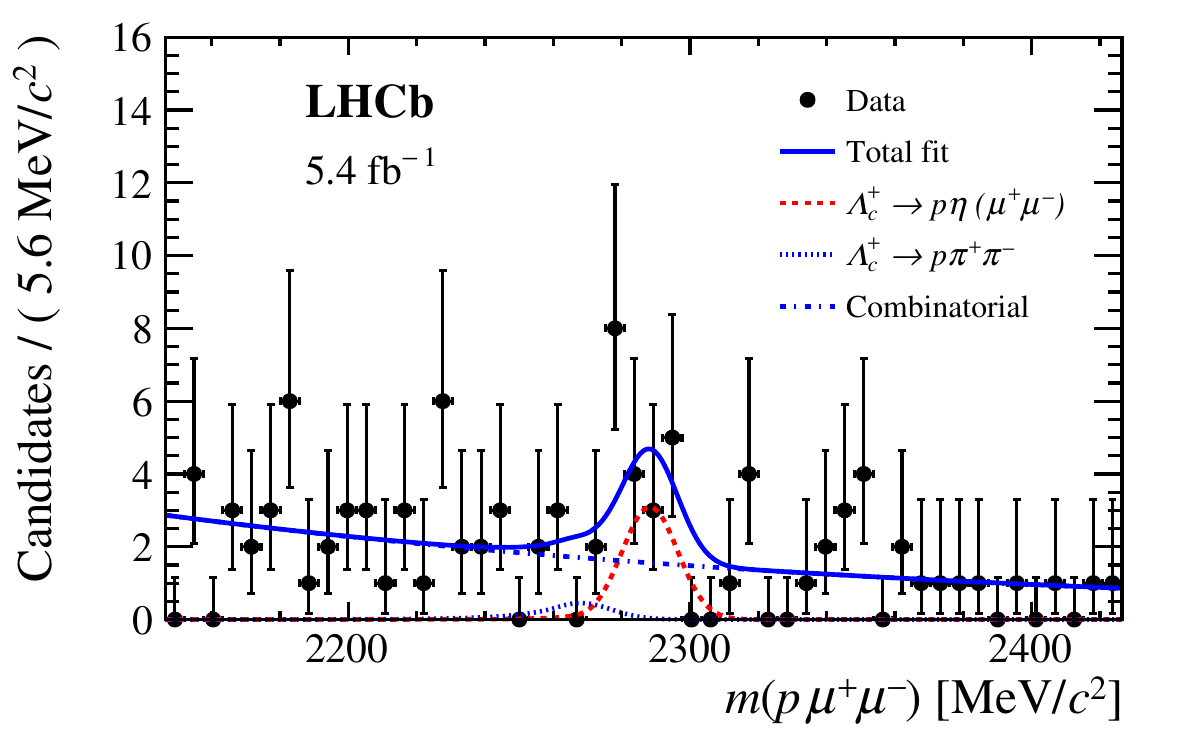}
     \vspace*{-0.6cm}
  \end{center}
  \caption{
    \small 
      Distributions of $p\mu^+\mu^-$ invariant mass for (top left) $\phi$, (top right) $\omega$, (bottom left) $\rho$  and (bottom right) $\eta$ regions. The results of the fit are overlaid.
   }
   \label{fig:massfitresonant}
\end{figure}

The results of the fit to the data for the \emph{signal} region and for the \lowq and \highq regions are presented in Fig.~\ref{fig:massfitsignal}. 
The results of the fit for the normalisation mode and other regions dominated by resonances 
are presented in Fig.~\ref{fig:massfitresonant}. The yields and the \Lc signal significance for the six \mumum regions are reported in Table~\ref{tab:FitResultsInQ}, where the significance is determined using Wilks' theorem~\cite{Wilks:1938dza}. 
For the signal regions the upper limits for the corresponding branching fractions are determined.
The residual contribution from resonant decays into signal regions is estimated from simulation neglecting interference effects. For the \highq region, where the significance is the highest, only 2.8 events are attributed to \LcTopphi decays while for the \lowq region the expected contribution is negligible.
\begin{table}[b]
	\caption{
		Results of the fit to the \pmumu invariant-mass distribution for various \mumum regions. The significance of the \LcTopmumu component is also given.
	}
	\begin{center}\begin{tabular}{ccccc}
			\hline
	Region   & \LcTopmumu    &  \LcToppipi & Combinatorial   & Significance \\
	         & yield   &  yield  & yield  & \LcTopmumu \\
\hline
$\signal$ & $ \phantom{0}18 \pm 10 $           & $ \phantom{0,}3 \pm 7\phantom{0,} $ & $ 681 \pm 28 $ & $ 2.0\sigma$ \\
$\lowq$   & $ \phantom{00}1 \pm 5\phantom{0} $ & $ \phantom{0,}4 \pm 4\phantom{0,} $ & $ 241 \pm 17 $ & $ 0.3\sigma$ \\
$\highq$  & $ \phantom{0}21 \pm 8\phantom{0} $ & $ \phantom{0,}4 \pm 4\phantom{0,} $ & $ 432 \pm 22 $ & $ 2.8\sigma$ \\
$\eta$    & $ \phantom{0}12 \pm 5\phantom{0} $ & $ 2.2 \pm 1.6 $                     & $ \phantom{0}84 \pm 10 $ & $ 3.0\sigma$ \\
$\rho$    & $ \phantom{0}43 \pm 10 $           & $ \phantom{,}20 \pm 6\phantom{0,} $ & $ 382 \pm 22 $ & $ 5.6\sigma$ \\
$\omega$  & $ \phantom{0}81 \pm 10 $           & $ 4.8 \pm 2.1 $                     & $ 101 \pm 11 $ & $ >7\sigma$ \\
$\phi$    & $ 423 \pm 22 $                     & $ 3.8 \pm 2.4 $                     & $ 173 \pm 15 $ & $ >7\sigma$ \\
\hline
		\end{tabular}
	\end{center}
	\label{tab:FitResultsInQ}
\end{table}

\section{Systematic uncertainties}
%
The systematic uncertainties taken into account in the determination of the upper limit on the branching fraction are summarised in Table~\ref{tab:sigsyst}. The two main contributions are the statistical uncertainty of the simulation samples used to determine the efficiency ratio and the uncertainty on the yield of the normalisation mode. The latter is dominated by the statistical uncertainty on the yield returned from the fit.
Other sources of systematic uncertainty on the efficiency ratio are associated with the following:
\begin{table}[t]
  \caption{
    Summary of systematic uncertainties for the signal regions and for the $\eta$,   $\rho$  and $\omega$ resonant regions.}
\setlength{\tabcolsep}{10pt}
\renewcommand{\arraystretch}{1.1}
\begin{center}\begin{tabular}{lcccccc}
    \hline
 Uncertainty source [\%] & \textit{signal}   & \lowq  & \highq & $\eta$   & $\rho$  & $\omega$   \\
 \hline
  Normalisation channel          & 5.3   & 5.3   & 5.3   & 5.3   & 5.3  & 5.3     \\
  Efficiency ratio (\rm stat.)   & 5.7   & 8.2   & 6.8   & 12.9  & 5.3  & 4.6     \\
  Efficiency ratio (\rm syst.)   & 1.7   & 1.7   & 1.7   & 1.7   & 1.7  & 1.7     \\
  Shape of signal                & 1.6   & 3.4   & 1.1   & 0.6   & 1.7  & 1.2     \\
  Shape of \LcToppipi            & 0.1   & 3.2   & 0.2   & 0.4   & 5.3  & 1.2     \\
  Shape of combinatorial         & 0.1   & 0.3   & 0.2   & 0.1   & 0.0  & 0.0     \\
  \LcToppipi decay model         & 0.1   & 0.1   & 0.1   & 0.1   & 0.1  & 0.1     \\
  Fit bias                       & 0.1   & 0.1   & 0.2   & 0.3   & 0.1  & 0.1     \\
  \hline
  Total                          & 8.1   & 11.0  & 8.9  & 14.1  & 9.5  & 7.4   \\
  \hline
  \end{tabular}\end{center}
\label{tab:sigsyst}
\end{table}

\begin{itemize} 
\item
The level of agreement between data and simulation.
The maximum difference between ratios of efficiencies determined for a few variants of the baseline procedure of weighting simulation to match the data is taken as a systematic uncertainty.
\item
The efficiency determination from simulation.
The difference in the efficiency obtained including or excluding \Lc candidates for which one or more decay products are not correctly matched to the generated particles, is taken as a systematic uncertainty. The procedure is applied consistently for the signal and normalisation modes.
\item
The correction of PID variables in simulation.
The difference of the efficiency ratio for the baseline and an alternative parameterisation of the PID calibration procedure~\cite{LHCb-DP-2018-001} is taken as a systematic uncertainty.
\end{itemize} 
These three sources added in quadrature amount to a relative uncertainty of 1.7\%.

The systematic uncertainties related to the fit model are estimated using pseudoexperiments. An alternative model for each component is used to generate pseudodata, which are then fitted using the baseline model described in Sec.~\ref{sec:fit}. 
A Johnson function~\cite{JohnsonNL} is used as an alternative shape for the signal, 
while the \LcToppipi background is parameterised as the sum of a Johnson and a Gaussian function, and a linear function is used as an alternative model for the combinatorial background. 
The difference between the generated signal yield and the average fitted yield is taken as systematic uncertainty.
In addition, the uncertainty related to the \LcToppipi decay model is determined by applying different corrections to the \LcToppipi simulated sample and determining an alternative invariant-mass shape. The bias due to maximum-likelihood fit is determined from the same data model used for pseudodata generation and fitting. The signal yield difference between the generated value and the average of fitted values from the pseudoexperiments is taken as a systematic uncertainty.

The systematic uncertainties associated with the ratio of branching fractions in the resonant regions are determined in similar way as for the signal regions and are presented in Table~\ref{tab:sigsyst}.

\section{Results}
%
\subsection{Signal regions}
\begin{figure}
  \begin{center}
    \includegraphics[width=0.6\textwidth]{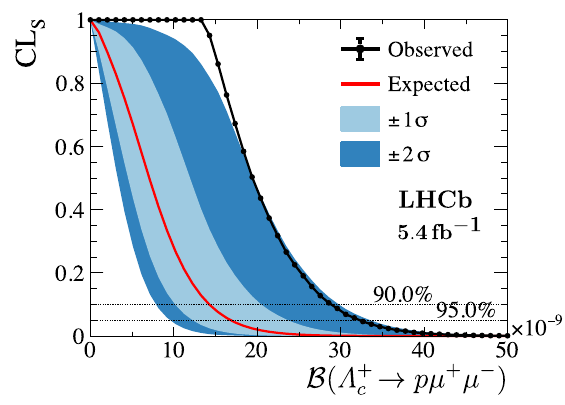}
  \end{center}
  \caption{
Observed and expected (background only hypothesis) limits for $\Lambda_c^+ \to p \mu^+ \mu^-$  branching fraction.}
  \label{fig:clsbf}
\end{figure}

%
No significant contribution of \LcTopmumu decays is observed in either of the two signal regions. Upper limits on the branching fractions and their ratios are determined using the \CLs method implemented in the G\textsc{amma}C\textsc{ombo} package~\cite{GammaCombo}.
The systematic uncertainties are included in the determination of \CLs. 
The following upper limit in the \emph{signal} region is determined to be   
\begin{align*}                                                           \frac{ \BRof{\LcTopmumu} }{\BRof{\Lcpphino}\BRof{\phimumu}  }  < 0.09~(0.10)~~{\rm at}~~90\%~(95\%)~{\rm CL}.
\end{align*}                                    
Using the values of the branching fractions for \Lcpphino and \phimumu decays from Ref.~\cite{PDG2022} and including their uncertainties in the \CLs determination, an upper limit on the branching fraction is determined to be 
\begin{align*} 
\BRof{\LcTopmumu} < 2.9~(3.2) \times 10^{-8}~~~~{\rm at}~~90\%~(95\%)~{\rm CL}.
 \end{align*} 
The \CLs as a function of branching fraction is shown in Fig.~\ref{fig:clsbf}.
The limits are determined for \lowq and \highq regions separately with the same assumptions as for the \textit{signal} region.  
The following upper limits are determined: 
\begin{align*}                                           
\frac{ \BRof{\LcTopmumu} }{  \BRof{\Lcpphino}\BRof{\phimumu}  }  < 0.029~(0.034)~~{\rm at}~~90\%~(95\%)~{\rm CL}~(\lowq), \\[3pt]
\frac{ \BRof{\LcTopmumu} }{  \BRof{\Lcpphino}\BRof{\phimumu}  }  < 0.094~(0.10)~~{\rm at}~~90\%~(95\%)~{\rm CL}~(\highq), \\[6pt]
\vspace{0.5cm}
\BRof{\LcTopmumu} < 0.93~(1.1) \times 10^{-8}~~~~{\rm at}~~90\%~(95\%)~{\rm CL}~(\lowq), \\[3pt]
\BRof{\LcTopmumu} < 3.0~(3.3) \times 10^{-8}~~~~{\rm at}~~90\%~(95\%)~{\rm CL}~(\highq).
\end{align*} 
The upper limit placed using data from the \highq region is driven by the signal yield whose significance is close to but lower than 3$\sigma$, while in the \lowq region the invariant-mass distribution is fully compatible with no signal.

Assuming a decay model with final state particles equally distributed in the phase space, the values of upper limits extrapolated from the \textit{signal} region to the full phase space are determined to be
\begin{align*}                                                           \frac{ \BRof{\LcTopmumu} }{\BRof{\Lcpphino}\BRof{\phimumu}  }  < 0.23~(0.25)~~{\rm at}~~90\%~(95\%)~{\rm CL}, \\[3pt]
\BRof{\LcTopmumu} < 7.3~(8.2) \times 10^{-8}~~~~{\rm at}~~90\%~(95\%)~{\rm CL}.
 \end{align*} 
In spite of the enlarged data sample, the extrapolated upper limit on the branching fraction is only slightly better than that determined for the 2011--2012 data~\cite{LHCb-PAPER-2017-039}, \mbox{$7.7~(9.6) \times 10^{-8}~{\rm at}~~90\%~(95\%)~{\rm CL}$}. 
This relatively small improvement is driven mainly by the 2.8$\sigma$ effect observed in the \highq region in this analysis.

\subsection{Resonant regions}
%
Excluding the normalisation channel, the significance of the \Lc signal exceeds the 5$\sigma$ threshold in the $\rho$ and $\omega$ regions while  
the significance for the \LcTopeta decay is at $3\sigma$. 
The corresponding branching fractions are determined using 
Eqs.~\ref{eq:normalisation},\ref{eq:alpha} 
where the decay parameters for the nonresonant decays are replaced by those for the resonant decays. 
It has to be emphasised that disentangling the resonant components to account for interference effects has not been attempted. 
The efficiency for a given \mumum region is determined using simulation samples of the corresponding resonant decay, thus neglecting possible contributions from other resonances.
Under this assumption, the ratio of branching fractions with respect to the normalisation channel are determined for the $\omega$, $\rho$ and $\eta$ regions to be

\begin{align*}
\frac{\BRof{\LcTopomega}}{\BRof{\LcTopphi}} = 0.240 \pm 0.030~(\rm stat.) \pm 0.018~(\rm syst.), \\[3pt]
\frac{\BRof{\LcToprho}}{\BRof{\LcTopphi}} = 0.229 \pm 0.051~(\rm stat.) \pm 0.022~(\rm syst.), \\[3pt]
\frac{\BRof{\LcTopeta}}{\BRof{\LcTopphi}} = 0.032 \pm 0.013~(\rm stat.) \pm 0.004~(\rm syst.).
\end{align*}
 The corresponding branching fractions are determined to be
\begin{align*}
\BRof{\Lcpomegano} = (9.82 \pm 1.23~(\rm stat.) \pm 0.73~(\rm syst.) \pm 2.79~(ext.) )\times 10^{-4}, \\[3pt]
\BRof{\Lcprhono} = (1.52 \pm 0.34~(\rm stat.) \pm 0.14~(\rm syst.) \pm 0.24~(ext.) )\times 10^{-3}, \\[3pt]
\BRof{\Lcpetano} = (1.67 \pm 0.69~(\rm stat.) \pm 0.23~(\rm syst.) \pm 0.34~(ext.) )\times 10^{-3},
\end{align*}
where the second uncertainties are induced by the ratio of branching fractions presented in Table~\ref{tab:sigsyst} and the third uncertainties are due to the following branching fractions: $\BRof{\Lcpphino}$, $\BRof{\phimumu}$, $\BRof{\omegamumu}$, $\BRof{\rhomumu}$ and $\BRof{\etamumu}$~\cite{PDG2022}, which were used as external inputs. 

\section{Summary}
A search for the nonresonant \LcTopmumu decay is presented using data samples of proton-proton collisions collected by the LHCb experiment in 2016--2018, corresponding to an integrated luminosity of $5.4\invfb$. 
In the dimuon invariant-mass regions where the contributions of resonant decays is subdominant, {\it i.e.} below the $\eta$ and above the $\phi$ resonances, the following upper limits for the branching fractions are derived:  \mbox{$\BRof{\LcTopmumu} < 0.93~(1.1) \times 10^{-8}$}  at \mbox{90\%~(95\%) CL} and  \mbox{$\BRof{\LcTopmumu} < 3.0~(3.3) \times 10^{-8}$}  at \mbox{90\%~(95\%) CL}, respectively where \mbox{$\LcTopphi$} as a normalisation channel. Combining these two regions, one finds \mbox{$\BRof{\LcTopmumu} < 2.9~(3.2) \times 10^{-8}$} at \mbox{90\%~(95\%) CL}.
For the reasons described in the introduction, the results for the short-distance contribution presented in this paper should not be combined with the results published for 2011--2012 data~\cite{LHCb-PAPER-2017-039}.

In a narrow region around the $\omega$ resonance, a \Lc signal is observed with a significance of more than 7$\sigma$, corresponding to a branching fraction  \mbox{$(9.82 \pm 1.23~(\rm stat.) \pm 0.73~(\rm syst.) \pm 2.79~(ext.) )\times 10^{-4}$}, compatible with and more precise than the result of the previous analysis for the data collected in 2011--2012~\cite{LHCb-PAPER-2017-039}.

Excluding the region around the $\omega$ meson, a signal in the $\rho$ region is found with a significance of 5.6$\sigma$. The corresponding branching fraction determined with respect to the normalisation channel is measured to be  \mbox{$(1.52 \pm 0.34~(\rm stat.) \pm 0.14~(\rm syst.) \pm 0.24~(ext.) )\times 10^{-3}$}, giving strong evidence for the $\Lcprhono$ decay, qualified by the fact that interference effects are not accounted for.
For the $\eta$ region the branching fraction was measured to be \mbox{$(1.67 \pm 0.69~(\rm stat.) \pm 0.23~(\rm syst.) \pm 0.34~(ext.) )\times 10^{-3}$} with a significance of 3.0$\sigma$. The value is compatible with the world average~\cite{PDG2022} but less precise.

\section*{Acknowledgements}
%
%
\noindent We express our gratitude to our colleagues in the CERN
accelerator departments for the excellent performance of the LHC. We
thank the technical and administrative staff at the LHCb
institutes.
We acknowledge support from CERN and from the national agencies:
CAPES, CNPq, FAPERJ and FINEP (Brazil); 
MOST and NSFC (China); 
CNRS/IN2P3 (France); 
BMBF, DFG and MPG (Germany); 
INFN (Italy); 
NWO (Netherlands); 
MNiSW and NCN (Poland); 
MCID/IFA (Romania); 
MICIU and AEI (Spain);
SNSF and SER (Switzerland); 
NASU (Ukraine); 
STFC (United Kingdom); 
DOE NP and NSF (USA).
We acknowledge the computing resources that are provided by CERN, IN2P3
(France), KIT and DESY (Germany), INFN (Italy), SURF (Netherlands),
PIC (Spain), GridPP (United Kingdom), 
CSCS (Switzerland), IFIN-HH (Romania), CBPF (Brazil),
and Polish WLCG (Poland).
We are indebted to the communities behind the multiple open-source
software packages on which we depend.
Individual groups or members have received support from
ARC and ARDC (Australia);
Key Research Program of Frontier Sciences of CAS, CAS PIFI, CAS CCEPP, 
Fundamental Research Funds for the Central Universities, 
and Sci. \& Tech. Program of Guangzhou (China);
Minciencias (Colombia);
EPLANET, Marie Sk\l{}odowska-Curie Actions, ERC and NextGenerationEU (European Union);
A*MIDEX, ANR, IPhU and Labex P2IO, and R\'{e}gion Auvergne-Rh\^{o}ne-Alpes (France);
AvH Foundation (Germany);
ICSC (Italy); 
Severo Ochoa and Mar\'ia de Maeztu Units of Excellence, GVA, XuntaGal, GENCAT, InTalent-Inditex and Prog. ~Atracci\'on Talento CM (Spain);
SRC (Sweden);
the Leverhulme Trust, the Royal Society
 and UKRI (United Kingdom).

\clearpage



\addcontentsline{toc}{section}{References}
\bibliographystyle{LHCb}
\bibliography{main,standard,LHCb-PAPER,LHCb-CONF,LHCb-DP,LHCb-TDR,myref}

\ifx\mcitethebibliography\mciteundefinedmacro
\PackageError{LHCb.bst}{mciteplus.sty has not been loaded}
{This bibstyle requires the use of the mciteplus package.}\fi
\providecommand{\href}[2]{#2}
\begin{mcitethebibliography}{10}
\mciteSetBstSublistMode{n}
\mciteSetBstMaxWidthForm{subitem}{\alph{mcitesubitemcount})}
\mciteSetBstSublistLabelBeginEnd{\mcitemaxwidthsubitemform\space}
{\relax}{\relax}

\bibitem{GIM}
S.~Glashow, J.~Iliopoulos, and L.~Maiani,
  \ifthenelse{\boolean{articletitles}}{\emph{{Weak interactions with
  lepton-hadron symmetry}},
  }{}\href{https://doi.org/10.1103/PhysRevD.2.1285}{Phys.\ Rev.\  \textbf{D2}
  (1970) 1285}\relax
\mciteBstWouldAddEndPuncttrue
\mciteSetBstMidEndSepPunct{\mcitedefaultmidpunct}
{\mcitedefaultendpunct}{\mcitedefaultseppunct}\relax
\EndOfBibitem
\bibitem{Meinel:2017ggx}
S.~Meinel, \ifthenelse{\boolean{articletitles}}{\emph{{$\Lambda_c \to N$ form
  factors from lattice QCD and phenomenology of $\Lambda_c \to n \ell^+
  \nu_\ell$ and $\Lambda_c \to p \mu^+ \mu^-$ decays}},
  }{}\href{https://doi.org/10.1103/PhysRevD.97.034511}{Phys.\ Rev.\
  \textbf{D97} (2018) 034511},
  \href{http://arxiv.org/abs/1712.05783}{{\normalfont\ttfamily
  arXiv:1712.05783}}\relax
\mciteBstWouldAddEndPuncttrue
\mciteSetBstMidEndSepPunct{\mcitedefaultmidpunct}
{\mcitedefaultendpunct}{\mcitedefaultseppunct}\relax
\EndOfBibitem
\bibitem{GoltzRare2021}
M.~Golz, G.~Hiller, and T.~Magorsch,
  \ifthenelse{\boolean{articletitles}}{\emph{{Probing for new physics with rare
  charm baryon ($\Lambda_c$, $\Xi_c$, $\Omega_c$) decays}},
  }{}\href{https://doi.org/https://doi.org/10.1007/JHEP09(2021)208}{JHEP
  \textbf{09} (2021) 208},
  \href{http://arxiv.org/abs/2107.13010}{{\normalfont\ttfamily
  arXiv:2107.13010}}\relax
\mciteBstWouldAddEndPuncttrue
\mciteSetBstMidEndSepPunct{\mcitedefaultmidpunct}
{\mcitedefaultendpunct}{\mcitedefaultseppunct}\relax
\EndOfBibitem
\bibitem{CHARM_RARE_FAJFER}
S.~Fajfer.\, S.~Prelovsek, and P.~Singer,
  \ifthenelse{\boolean{articletitles}}{\emph{{Rare charm meson decays $D \to P
  l^+l^-$ and $c \to ul^+l^-$ in the standard model and the minimal
  supersymmetric standard model}},
  }{}\href{https://doi.org/10.1103/PhysRevD.64.114009}{Phys.\ Rev.\
  \textbf{D64} (2001) 114009},
  \href{http://arxiv.org/abs/hep-ph/0106333}{{\normalfont\ttfamily
  arXiv:hep-ph/0106333}}\relax
\mciteBstWouldAddEndPuncttrue
\mciteSetBstMidEndSepPunct{\mcitedefaultmidpunct}
{\mcitedefaultendpunct}{\mcitedefaultseppunct}\relax
\EndOfBibitem
\bibitem{CHARM_RARE_BURDMAN}
G.~Burdman, E.~Golowich, J.~Hewett, and S.~Pakvasa,
  \ifthenelse{\boolean{articletitles}}{\emph{{Rare charm decays in the standard
  model and beyond}},
  }{}\href{https://doi.org/10.1103/PhysRevD.66.014009}{Phys.\ Rev.\
  \textbf{D66} (2002) 014009}\relax
\mciteBstWouldAddEndPuncttrue
\mciteSetBstMidEndSepPunct{\mcitedefaultmidpunct}
{\mcitedefaultendpunct}{\mcitedefaultseppunct}\relax
\EndOfBibitem
\bibitem{LHCb-PAPER-2020-041}
LHCb collaboration, R.~Aaij {\em et~al.},
  \ifthenelse{\boolean{articletitles}}{\emph{{Angular analysis of the $B^{+}\to
  K^{\ast+}\mu^+\mu^-$ decay}},
  }{}\href{https://doi.org/10.1103/PhysRevLett.126.161802}{Phys.\ Rev.\ Lett.\
  \textbf{126} (2021) 161802},
  \href{http://arxiv.org/abs/2012.13241}{{\normalfont\ttfamily
  arXiv:2012.13241}}\relax
\mciteBstWouldAddEndPuncttrue
\mciteSetBstMidEndSepPunct{\mcitedefaultmidpunct}
{\mcitedefaultendpunct}{\mcitedefaultseppunct}\relax
\EndOfBibitem
\bibitem{LHCb-PAPER-2021-035}
LHCb collaboration, R.~Aaij {\em et~al.},
  \ifthenelse{\boolean{articletitles}}{\emph{{Angular analysis of $D^0 \to
  \pi^+ \pi ^- \mu^+ \mu^-$ and $D^0 \to K^+ K ^- \mu^+ \mu^-$ decays and
  search for \CP violation}},
  }{}\href{https://doi.org/10.1103/PhysRevLett.128.221801}{Phys.\ Rev.\ Lett.\
  \textbf{128} (2022) 221801},
  \href{http://arxiv.org/abs/2111.03327}{{\normalfont\ttfamily
  arXiv:2111.03327}}\relax
\mciteBstWouldAddEndPuncttrue
\mciteSetBstMidEndSepPunct{\mcitedefaultmidpunct}
{\mcitedefaultendpunct}{\mcitedefaultseppunct}\relax
\EndOfBibitem
\bibitem{MitzelRare2021}
H.~Gisbert, M.~Golz, and D.~Mitzel,
  \ifthenelse{\boolean{articletitles}}{\emph{{Theoretical and experimental
  status of rare charm decays}},
  }{}\href{https://doi.org/http://dx.doi.org/10.1142/S0217732321300020}{Mod.\
  Phys.\ Lett.\  \textbf{A36} (2021) 2130002},
  \href{http://arxiv.org/abs/2011.09478}{{\normalfont\ttfamily
  arXiv:2011.09478}}\relax
\mciteBstWouldAddEndPuncttrue
\mciteSetBstMidEndSepPunct{\mcitedefaultmidpunct}
{\mcitedefaultendpunct}{\mcitedefaultseppunct}\relax
\EndOfBibitem
\bibitem{LHCb-PAPER-2017-039}
LHCb collaboration, R.~Aaij {\em et~al.},
  \ifthenelse{\boolean{articletitles}}{\emph{{Search for the rare decay
  \mbox{\decay{\Lc}{p\mumu}}}},
  }{}\href{https://doi.org/10.1103/PhysRevD.97.091101}{Phys.\ Rev.\
  \textbf{D97} (2018) 091101},
  \href{http://arxiv.org/abs/1712.07938}{{\normalfont\ttfamily
  arXiv:1712.07938}}\relax
\mciteBstWouldAddEndPuncttrue
\mciteSetBstMidEndSepPunct{\mcitedefaultmidpunct}
{\mcitedefaultendpunct}{\mcitedefaultseppunct}\relax
\EndOfBibitem
\bibitem{BABAR}
BABAR collaboration, J.~P. Lees {\em et~al.},
  \ifthenelse{\boolean{articletitles}}{\emph{{Searches for rare or forbidden
  semileptonic charm decays}},
  }{}\href{https://doi.org/10.1103/PhysRevD.84.072006}{Phys.\ Rev.\
  \textbf{D84} (2011) 072006},
  \href{http://arxiv.org/abs/1107.4465}{{\normalfont\ttfamily
  arXiv:1107.4465}}\relax
\mciteBstWouldAddEndPuncttrue
\mciteSetBstMidEndSepPunct{\mcitedefaultmidpunct}
{\mcitedefaultendpunct}{\mcitedefaultseppunct}\relax
\EndOfBibitem
\bibitem{CLs}
A.~L. Read, \ifthenelse{\boolean{articletitles}}{\emph{{Presentation of search
  results: The CL$_{\rm s}$ technique}},
  }{}\href{https://doi.org/10.1088/0954-3899/28/10/313}{J.\ Phys.\
  \textbf{G28} (2002) 2693}\relax
\mciteBstWouldAddEndPuncttrue
\mciteSetBstMidEndSepPunct{\mcitedefaultmidpunct}
{\mcitedefaultendpunct}{\mcitedefaultseppunct}\relax
\EndOfBibitem
\bibitem{LHCb-DP-2008-001}
LHCb collaboration, A.~A. Alves~Jr.\ {\em et~al.},
  \ifthenelse{\boolean{articletitles}}{\emph{{The \lhcb detector at the LHC}},
  }{}\href{https://doi.org/10.1088/1748-0221/3/08/S08005}{JINST \textbf{3}
  (2008) S08005}\relax
\mciteBstWouldAddEndPuncttrue
\mciteSetBstMidEndSepPunct{\mcitedefaultmidpunct}
{\mcitedefaultendpunct}{\mcitedefaultseppunct}\relax
\EndOfBibitem
\bibitem{LHCb-DP-2014-002}
LHCb collaboration, R.~Aaij {\em et~al.},
  \ifthenelse{\boolean{articletitles}}{\emph{{LHCb detector performance}},
  }{}\href{https://doi.org/10.1142/S0217751X15300227}{Int.\ J.\ Mod.\ Phys.\
  \textbf{A30} (2015) 1530022},
  \href{http://arxiv.org/abs/1412.6352}{{\normalfont\ttfamily
  arXiv:1412.6352}}\relax
\mciteBstWouldAddEndPuncttrue
\mciteSetBstMidEndSepPunct{\mcitedefaultmidpunct}
{\mcitedefaultendpunct}{\mcitedefaultseppunct}\relax
\EndOfBibitem
\bibitem{LHCb-DP-2014-001}
R.~Aaij {\em et~al.}, \ifthenelse{\boolean{articletitles}}{\emph{{Performance
  of the LHCb Vertex Locator}},
  }{}\href{https://doi.org/10.1088/1748-0221/9/09/P09007}{JINST \textbf{9}
  (2014) P09007}, \href{http://arxiv.org/abs/1405.7808}{{\normalfont\ttfamily
  arXiv:1405.7808}}\relax
\mciteBstWouldAddEndPuncttrue
\mciteSetBstMidEndSepPunct{\mcitedefaultmidpunct}
{\mcitedefaultendpunct}{\mcitedefaultseppunct}\relax
\EndOfBibitem
\bibitem{LHCb-DP-2017-001}
P.~d'Argent {\em et~al.}, \ifthenelse{\boolean{articletitles}}{\emph{{Improved
  performance of the LHCb Outer Tracker in LHC Run 2}},
  }{}\href{https://doi.org/10.1088/1748-0221/12/11/P11016}{JINST \textbf{12}
  (2017) P11016}, \href{http://arxiv.org/abs/1708.00819}{{\normalfont\ttfamily
  arXiv:1708.00819}}\relax
\mciteBstWouldAddEndPuncttrue
\mciteSetBstMidEndSepPunct{\mcitedefaultmidpunct}
{\mcitedefaultendpunct}{\mcitedefaultseppunct}\relax
\EndOfBibitem
\bibitem{LHCb-DP-2012-003}
M.~Adinolfi {\em et~al.},
  \ifthenelse{\boolean{articletitles}}{\emph{{Performance of the \lhcb RICH
  detector at the LHC}},
  }{}\href{https://doi.org/10.1140/epjc/s10052-013-2431-9}{Eur.\ Phys.\ J.\
  \textbf{C73} (2013) 2431},
  \href{http://arxiv.org/abs/1211.6759}{{\normalfont\ttfamily
  arXiv:1211.6759}}\relax
\mciteBstWouldAddEndPuncttrue
\mciteSetBstMidEndSepPunct{\mcitedefaultmidpunct}
{\mcitedefaultendpunct}{\mcitedefaultseppunct}\relax
\EndOfBibitem
\bibitem{LHCb-DP-2012-002}
A.~A. Alves~Jr.\ {\em et~al.},
  \ifthenelse{\boolean{articletitles}}{\emph{{Performance of the LHCb muon
  system}}, }{}\href{https://doi.org/10.1088/1748-0221/8/02/P02022}{JINST
  \textbf{8} (2013) P02022},
  \href{http://arxiv.org/abs/1211.1346}{{\normalfont\ttfamily
  arXiv:1211.1346}}\relax
\mciteBstWouldAddEndPuncttrue
\mciteSetBstMidEndSepPunct{\mcitedefaultmidpunct}
{\mcitedefaultendpunct}{\mcitedefaultseppunct}\relax
\EndOfBibitem
\bibitem{LHCb-DP-2012-004}
R.~Aaij {\em et~al.}, \ifthenelse{\boolean{articletitles}}{\emph{{The \lhcb
  trigger and its performance in 2011}},
  }{}\href{https://doi.org/10.1088/1748-0221/8/04/P04022}{JINST \textbf{8}
  (2013) P04022}, \href{http://arxiv.org/abs/1211.3055}{{\normalfont\ttfamily
  arXiv:1211.3055}}\relax
\mciteBstWouldAddEndPuncttrue
\mciteSetBstMidEndSepPunct{\mcitedefaultmidpunct}
{\mcitedefaultendpunct}{\mcitedefaultseppunct}\relax
\EndOfBibitem
\bibitem{Sjostrand:2007gs}
T.~Sj\"{o}strand, S.~Mrenna, and P.~Skands,
  \ifthenelse{\boolean{articletitles}}{\emph{{A brief introduction to PYTHIA
  8.1}}, }{}\href{https://doi.org/10.1016/j.cpc.2008.01.036}{Comput.\ Phys.\
  Commun.\  \textbf{178} (2008) 852},
  \href{http://arxiv.org/abs/0710.3820}{{\normalfont\ttfamily
  arXiv:0710.3820}}\relax
\mciteBstWouldAddEndPuncttrue
\mciteSetBstMidEndSepPunct{\mcitedefaultmidpunct}
{\mcitedefaultendpunct}{\mcitedefaultseppunct}\relax
\EndOfBibitem
\bibitem{Sjostrand:2006za}
T.~Sj\"{o}strand, S.~Mrenna, and P.~Skands,
  \ifthenelse{\boolean{articletitles}}{\emph{{PYTHIA 6.4 physics and manual}},
  }{}\href{https://doi.org/10.1088/1126-6708/2006/05/026}{JHEP \textbf{05}
  (2006) 026}, \href{http://arxiv.org/abs/hep-ph/0603175}{{\normalfont\ttfamily
  arXiv:hep-ph/0603175}}\relax
\mciteBstWouldAddEndPuncttrue
\mciteSetBstMidEndSepPunct{\mcitedefaultmidpunct}
{\mcitedefaultendpunct}{\mcitedefaultseppunct}\relax
\EndOfBibitem
\bibitem{LHCb-PROC-2010-056}
I.~Belyaev {\em et~al.}, \ifthenelse{\boolean{articletitles}}{\emph{{Handling
  of the generation of primary events in Gauss, the LHCb simulation
  framework}}, }{}\href{https://doi.org/10.1088/1742-6596/331/3/032047}{J.\
  Phys.\ Conf.\ Ser.\  \textbf{331} (2011) 032047}\relax
\mciteBstWouldAddEndPuncttrue
\mciteSetBstMidEndSepPunct{\mcitedefaultmidpunct}
{\mcitedefaultendpunct}{\mcitedefaultseppunct}\relax
\EndOfBibitem
\bibitem{Lange:2001uf}
D.~J. Lange, \ifthenelse{\boolean{articletitles}}{\emph{{The EvtGen particle
  decay simulation package}},
  }{}\href{https://doi.org/10.1016/S0168-9002(01)00089-4}{Nucl.\ Instrum.\
  Meth.\  \textbf{A462} (2001) 152}\relax
\mciteBstWouldAddEndPuncttrue
\mciteSetBstMidEndSepPunct{\mcitedefaultmidpunct}
{\mcitedefaultendpunct}{\mcitedefaultseppunct}\relax
\EndOfBibitem
\bibitem{davidson2015photos}
N.~Davidson, T.~Przedzinski, and Z.~Was,
  \ifthenelse{\boolean{articletitles}}{\emph{{PHOTOS interface in C++:
  Technical and physics documentation}},
  }{}\href{https://doi.org/https://doi.org/10.1016/j.cpc.2015.09.013}{Comp.\
  Phys.\ Comm.\  \textbf{199} (2016) 86},
  \href{http://arxiv.org/abs/1011.0937}{{\normalfont\ttfamily
  arXiv:1011.0937}}\relax
\mciteBstWouldAddEndPuncttrue
\mciteSetBstMidEndSepPunct{\mcitedefaultmidpunct}
{\mcitedefaultendpunct}{\mcitedefaultseppunct}\relax
\EndOfBibitem
\bibitem{Allison:2006ve}
Geant4 collaboration, J.~Allison {\em et~al.},
  \ifthenelse{\boolean{articletitles}}{\emph{{Geant4 developments and
  applications}}, }{}\href{https://doi.org/10.1109/TNS.2006.869826}{IEEE
  Trans.\ Nucl.\ Sci.\  \textbf{53} (2006) 270}\relax
\mciteBstWouldAddEndPuncttrue
\mciteSetBstMidEndSepPunct{\mcitedefaultmidpunct}
{\mcitedefaultendpunct}{\mcitedefaultseppunct}\relax
\EndOfBibitem
\bibitem{Agostinelli:2002hh}
Geant4 collaboration, S.~Agostinelli {\em et~al.},
  \ifthenelse{\boolean{articletitles}}{\emph{{Geant4: A simulation toolkit}},
  }{}\href{https://doi.org/10.1016/S0168-9002(03)01368-8}{Nucl.\ Instrum.\
  Meth.\  \textbf{A506} (2003) 250}\relax
\mciteBstWouldAddEndPuncttrue
\mciteSetBstMidEndSepPunct{\mcitedefaultmidpunct}
{\mcitedefaultendpunct}{\mcitedefaultseppunct}\relax
\EndOfBibitem
\bibitem{LHCb-PROC-2011-006}
M.~Clemencic {\em et~al.}, \ifthenelse{\boolean{articletitles}}{\emph{{The
  \lhcb simulation application, Gauss: Design, evolution and experience}},
  }{}\href{https://doi.org/10.1088/1742-6596/331/3/032023}{J.\ Phys.\ Conf.\
  Ser.\  \textbf{331} (2011) 032023}\relax
\mciteBstWouldAddEndPuncttrue
\mciteSetBstMidEndSepPunct{\mcitedefaultmidpunct}
{\mcitedefaultendpunct}{\mcitedefaultseppunct}\relax
\EndOfBibitem
\bibitem{LHCb-DP-2018-004}
D.~M{\"u}ller, M.~Clemencic, G.~Corti, and M.~Gersabeck,
  \ifthenelse{\boolean{articletitles}}{\emph{{ReDecay: A novel approach to
  speed up the simulation at LHCb}},
  }{}\href{https://doi.org/10.1140/epjc/s10052-018-6469-6}{Eur.\ Phys.\ J.\
  \textbf{C78} (2018) 1009},
  \href{http://arxiv.org/abs/1810.10362}{{\normalfont\ttfamily
  arXiv:1810.10362}}\relax
\mciteBstWouldAddEndPuncttrue
\mciteSetBstMidEndSepPunct{\mcitedefaultmidpunct}
{\mcitedefaultendpunct}{\mcitedefaultseppunct}\relax
\EndOfBibitem
\bibitem{Breiman}
L.~Breiman, J.~H. Friedman, R.~A. Olshen, and C.~J. Stone, {\em Classification
  and regression trees}, Wadsworth international group, Belmont, California,
  USA, 1984\relax
\mciteBstWouldAddEndPuncttrue
\mciteSetBstMidEndSepPunct{\mcitedefaultmidpunct}
{\mcitedefaultendpunct}{\mcitedefaultseppunct}\relax
\EndOfBibitem
\bibitem{Hocker:2007ht}
H.~Voss, A.~Hoecker, J.~Stelzer, and F.~Tegenfeldt,
  \ifthenelse{\boolean{articletitles}}{\emph{{TMVA, Toolkit for Multivariate
  Data Analysis with ROOT}}, }{}\href{https://doi.org/10.22323/1.050.0040}{PoS
  \textbf{ACAT} (2007) 040}\relax
\mciteBstWouldAddEndPuncttrue
\mciteSetBstMidEndSepPunct{\mcitedefaultmidpunct}
{\mcitedefaultendpunct}{\mcitedefaultseppunct}\relax
\EndOfBibitem
\bibitem{TMVA4}
A.~Hoecker {\em et~al.}, \ifthenelse{\boolean{articletitles}}{\emph{{TMVA 4 --
  Toolkit for Multivariate Data Analysis with ROOT. Users Guide.}},
  }{}\href{http://arxiv.org/abs/physics/0703039}{{\normalfont\ttfamily
  arXiv:physics/0703039}}\relax
\mciteBstWouldAddEndPuncttrue
\mciteSetBstMidEndSepPunct{\mcitedefaultmidpunct}
{\mcitedefaultendpunct}{\mcitedefaultseppunct}\relax
\EndOfBibitem
\bibitem{fold}
A.~Blum, A.~Kalai, and J.~Langford,
  \ifthenelse{\boolean{articletitles}}{\emph{{Beating the hold-out: bounds for
  k-fold and progressive cross-validation}}, }{}COLT '99 Proceedings of the
  twelfth annual conference on Computational learning theory (1999) 203\relax
\mciteBstWouldAddEndPuncttrue
\mciteSetBstMidEndSepPunct{\mcitedefaultmidpunct}
{\mcitedefaultendpunct}{\mcitedefaultseppunct}\relax
\EndOfBibitem
\bibitem{LHCb-DP-2018-001}
R.~Aaij {\em et~al.}, \ifthenelse{\boolean{articletitles}}{\emph{{Selection and
  processing of calibration samples to measure the particle identification
  performance of the LHCb experiment in Run 2}},
  }{}\href{https://doi.org/10.1140/epjti/s40485-019-0050-z}{Eur.\ Phys.\ J.\
  Tech.\ Instr.\  \textbf{6} (2019) 1},
  \href{http://arxiv.org/abs/1803.00824}{{\normalfont\ttfamily
  arXiv:1803.00824}}\relax
\mciteBstWouldAddEndPuncttrue
\mciteSetBstMidEndSepPunct{\mcitedefaultmidpunct}
{\mcitedefaultendpunct}{\mcitedefaultseppunct}\relax
\EndOfBibitem
\bibitem{Skwarnicki:1986xj}
T.~Skwarnicki, {\em {A study of the radiative cascade transitions between the
  Upsilon-prime and Upsilon resonances}}, PhD thesis, Institute of Nuclear
  Physics, Krakow, 1986,
  {\href{http://inspirehep.net/record/230779/}{DESY-F31-86-02}}\relax
\mciteBstWouldAddEndPuncttrue
\mciteSetBstMidEndSepPunct{\mcitedefaultmidpunct}
{\mcitedefaultendpunct}{\mcitedefaultseppunct}\relax
\EndOfBibitem
\bibitem{LHCb-PAPER-2017-044}
LHCb collaboration, R.~Aaij {\em et~al.},
  \ifthenelse{\boolean{articletitles}}{\emph{{Search for \CP violation in
  \mbox{\decay{\Lc}{p \Km \Kp}} and \mbox{\decay{\Lc}{p\pim\pip}} decays}},
  }{}\href{https://doi.org/10.1007/JHEP03(2018)182}{JHEP \textbf{03} (2018)
  182}, \href{http://arxiv.org/abs/1712.07051}{{\normalfont\ttfamily
  arXiv:1712.07051}}\relax
\mciteBstWouldAddEndPuncttrue
\mciteSetBstMidEndSepPunct{\mcitedefaultmidpunct}
{\mcitedefaultendpunct}{\mcitedefaultseppunct}\relax
\EndOfBibitem
\bibitem{Wilks:1938dza}
S.~S. Wilks, \ifthenelse{\boolean{articletitles}}{\emph{{The large-sample
  distribution of the likelihood ratio for testing composite hypotheses}},
  }{}\href{https://doi.org/10.1214/aoms/1177732360}{Ann.\ Math.\ Stat.\
  \textbf{9} (1938) 60}\relax
\mciteBstWouldAddEndPuncttrue
\mciteSetBstMidEndSepPunct{\mcitedefaultmidpunct}
{\mcitedefaultendpunct}{\mcitedefaultseppunct}\relax
\EndOfBibitem
\bibitem{JohnsonNL}
N.~Johnson, \ifthenelse{\boolean{articletitles}}{\emph{{Systems of frequency
  curves generated by methods of translation}},
  }{}\href{https://doi.org/https://doi.org/10.1093/biomet/36.1-2.149}{Biometrica
  \textbf{36} (1949) 149}\relax
\mciteBstWouldAddEndPuncttrue
\mciteSetBstMidEndSepPunct{\mcitedefaultmidpunct}
{\mcitedefaultendpunct}{\mcitedefaultseppunct}\relax
\EndOfBibitem
\bibitem{GammaCombo}
M.~Kenzie {\em et~al.}, \ifthenelse{\boolean{articletitles}}{\emph{{GammaCombo:
  A statistical analysis framework for combining measurements, fitting datasets
  and producing confidence intervals}}, }{}
\newblock
  doi:~\href{https://doi.org/10.5281/zenodo.3371421}{10.5281/zenodo.3371421}\relax
\mciteBstWouldAddEndPuncttrue
\mciteSetBstMidEndSepPunct{\mcitedefaultmidpunct}
{\mcitedefaultendpunct}{\mcitedefaultseppunct}\relax
\EndOfBibitem
\bibitem{PDG2022}
Particle Data Group, R.~L. Workman {\em et~al.},
  \ifthenelse{\boolean{articletitles}}{\emph{{\href{http://pdg.lbl.gov/}{Review
  of particle physics}}}, }{}\href{https://doi.org/10.1093/ptep/ptac097}{Prog.\
  Theor.\ Exp.\ Phys.\  \textbf{2022} (2022) 083C01}\relax
\mciteBstWouldAddEndPuncttrue
\mciteSetBstMidEndSepPunct{\mcitedefaultmidpunct}
{\mcitedefaultendpunct}{\mcitedefaultseppunct}\relax
\EndOfBibitem
\end{mcitethebibliography}

\newpage
\centerline
{\large\bf LHCb collaboration}
\begin
{flushleft}
\small
R.~Aaij$^{36}$\lhcborcid{0000-0003-0533-1952},
A.S.W.~Abdelmotteleb$^{55}$\lhcborcid{0000-0001-7905-0542},
C.~Abellan~Beteta$^{49}$,
F.~Abudin{\'e}n$^{55}$\lhcborcid{0000-0002-6737-3528},
T.~Ackernley$^{59}$\lhcborcid{0000-0002-5951-3498},
A. A. ~Adefisoye$^{67}$\lhcborcid{0000-0003-2448-1550},
B.~Adeva$^{45}$\lhcborcid{0000-0001-9756-3712},
M.~Adinolfi$^{53}$\lhcborcid{0000-0002-1326-1264},
P.~Adlarson$^{79}$\lhcborcid{0000-0001-6280-3851},
C.~Agapopoulou$^{13}$\lhcborcid{0000-0002-2368-0147}
% ...

{\footnotesize \it

% ...
$^{13}$Universit{\'e} Paris-Saclay, CNRS/IN2P3, IJCLab, Orsay, France\\
% ...
$^{36}$Nikhef National Institute for Subatomic Physics, Amsterdam, Netherlands\\
% ...
$^{45}$Instituto Galego de F{\'\i}sica de Altas Enerx{\'\i}as (IGFAE), Universidade de Santiago de Compostela, Santiago de Compostela, Spain\\
% ...
$^{49}$Physik-Institut, Universit{\"a}t Z{\"u}rich, Z{\"u}rich, Switzerland\\
% ...
$^{53}$H.H. Wills Physics Laboratory, University of Bristol, Bristol, United Kingdom\\
% ...
$^{55}$Department of Physics, University of Warwick, Coventry, United Kingdom\\
% ...
$^{58}$School of Physics and Astronomy, University of Glasgow, Glasgow, United Kingdom\\
$^{59}$Oliver Lodge Laboratory, University of Liverpool, Liverpool, United Kingdom\\
% ...
$^{67}$Syracuse University, Syracuse, NY, United States\\
% ...
$^{79}$Department of Physics and Astronomy, Uppsala University, Uppsala, Sweden, associated to $^{58}$\\
% ...
}
\end{flushleft}

\end{document}